\documentclass[conference]{IEEEtran}
\usepackage{filecontents}
\usepackage{soul}

\usepackage[fleqn]{amsmath}
\usepackage{placeins}
\usepackage{float}
\usepackage{subcaption}
\usepackage{graphicx}
\usepackage{booktabs}
\graphicspath{{./figs}}
\usepackage{cite}

\ifCLASSINFOpdf
\else
\fi
\newlength \figwidth
\if@twocolumn
\else
\fi
\IEEEoverridecommandlockouts
\usepackage{atbegshi}
\AtBeginDocument{\AtBeginShipoutNext{\AtBeginShipoutDiscard}}
\begin{document}
%
\title{QoS Provisioning with Adaptive Backoff Algorithm for IEEE 802.11ac Under Multipacket Reception}


\author{\IEEEauthorblockN{Arun I B\IEEEauthorrefmark{0},
T. G. Venkatesh\IEEEauthorrefmark{0} and  Bhasker Dappuri\IEEEauthorrefmark{0} 
}
\IEEEauthorblockA{\IEEEauthorrefmark{0}Department of Electrical  Engineering, 
Indian  Institute of Technology Madras, Chennai 600036, India
\\ Email: arunib@gmail.com, tgvenky@ee.iitm.ac.in, bhasker.bvs@gmail.com}}

\thanks{In proc. of 9th International Symposium on Communication Systems, Networks \&  Digital Signal Processing (CSNDSP), 2014. \\
\copyright 2014 IEEE. Personal use of this material is permitted. Permission from IEEE must be obtained for all other uses, in any current or future media, including reprinting/republishing this material for advertising or promotional purposes, creating new collective works, for resale or redistribution to servers or lists, or reuse of any copyrighted component of this work in other works.
}


%
\maketitle
\begin{abstract}

Recent advances in physical layer communication techniques, enable receivers to decode multiple simultaneous transmissions. This technique is known as the multipacket reception (MPR). In this paper, we propose an enhancement to the IEEE 802.11ac EDCA protocol for channels supporting MPR for QoS provisioning. We show that in the case of MPR, in addition to $CW_{min}$, $CW_{max}$ and $AIFSN$, we can use two more parameters namely (i)threshold and (ii)counter decrement value, that can offer service differentiation. The performance evaluation of the different metrics of the proposed protocol (throughput, delay, and jitter) is carried out using extensive simulations.
\end{abstract}
\begin{IEEEkeywords}
Wireless LAN; 802.11ac; Quality of Service; Multipacket Reception; Medium Access Sublayer; Simulation
\end{IEEEkeywords}

%
\IEEEpeerreviewmaketitle

\section{Introduction}
\subsection{Background}
 With advances in physical (PHY) layer technologies such as multi user MIMO (MU-MIMO) \cite{Gesbert},\cite{Roy} it is now possible to receive or decode  several simultaneous transmissions in a wireless channel. This phenomenon is commonly referred to as the Multi-Packet Reception (MPR). In a MIMO system, multipacket reception is a consequence of having multiple spatial streams. In addition to MIMO, other physical layer technologies such as  Code Division Multiple Access, Orthogonal Frequency Division Multiple Access, Successive interference cancellation  also lead to multiuser detection(MUD) capabilities to the stations. For a detailed survey of the PHY layer technologies that enable MPR we refer the readers to \cite{Lu2012}.
 
 A related development in random matrix theory provides a means to estimate the number of packets involved in collision using a scheme called collision multiplicity estimation \cite{Benoit}.
 Going further, techniques have been proposed and even prototyped to sense the number of ongoing transmission in the channel \cite{Chan}\cite{Lin}. We refer to this carrier sensing capability as the enhanced carrier sensing. This enhanced carrier sensing capability is also known as multi-dimensional \cite{Lin} carrier sensing or MIMO \cite{Coviello2009} carrier sensing. 
 
 Traditional Medium Access Control (MAC)  protocols abstract the wireless channel using a collision channel model, wherein a packet can be received successfully at the receiver only if there is exactly one transmitter attempting to access the channel at a given slot. Simultaneous transmission of more than one packet leads to collision. However in multi access systems supporting MPR, nodes are allowed to transmit even when the channel is sensed to be busy. MAC protocols that are designed based on  the collision channel model, does not make use of this freedom offered by  systems supporting MPR. These MAC protocols underestimate channel capacity leading to inefficiency.  
 Thus traditional MAC protocols needs to be redesigned to leverage multiple access systems supporting multipacket reception and enhanced carrier sensing. MPR aware MAC protocols can not only improve the channel efficiency, but can also offer new freedom  in their design. 
 
New generation WLANs have to support data traffic of varying nature and applications with different constraints. While digital multimedia transmission in applications such as wireless display and gaming, have stringent QoS constraints, applications such as web browsing, sync and go file transfer are best effort in nature and therefore do not require QoS guarantee. MAC protocols for channels supporting MPR should also be able to offer service differentiation. 
 In the next section we briefly review the existing MPR-MAC protocols.
\subsection{Literature}
    One of the early works on MPR channel model is due to Ghez et al \cite{ghez1988}.    
Nagaraj et al. \cite{Nagaraj} provide an exact analysis for the throughput of pure Aloha for channels with MPR capability $K =2$, and an approximate expression for $K \geq2$. Arun and Venkatesh, based on an order statistics scheme have derived throughput of pure Aloha for channels with arbitrary MPR capability \cite{Arun}. 
MPR-MAC protocols based on enhanced carrier sensing have also been proposed. In \cite{dschan2004}, Chan and Berger have proposed a cross-layer solution for MPR known as cross layer CSMA (XL-CSMA) wherein station makes the decision to transmit based on information obtained from carrier sensing. Due to the wide spread popularity of the 802.11 Distributed Coordination Function (DCF) based WLANs, a number of MPR MAC protocols based on 802.11 DCF have been proposed.  
 Zheng et al.\cite{pxzheng2006} have proposed a protocol that modifies the packet structure of the CTS and ACK of 802.11 DCF to accommodate acknowledging of multiple stations. 
Later, Y.J. Zhang \cite{Junb} proposed  a multiround contention protocol in which several rounds of contention take place to select the winner before the data transmission. Barghi et al.\cite{Barghi2011} have proposed a MIMO based cross layer design in which some changes are made to the RTS-CTS signaling. Mahmood et al. \cite{Mahmood2010} have proposed a modification of DCF which obtains throughput gains by controlling the contention window size according to network loads. 

Recently Babich and Commiso \cite{fbabich2010} have proposed a generalization of 802.11 DCF to the MPR channels. In this threshold based protocol the backoff counter is frozen only when the number of ongoing transmissions in the channel is greater than a threshold. The ACK aware protocol of \cite{Mukhopad} is similar to the threshold based protocol in the sense that the node freezes its counter when the number of ongoing transmission is greater than a threshold. However the decrement process resumes when the channel is completely idle. In an earlier work we have proposed an adaptive backoff algorithm for the IEEE 802.11 DCF for MPR wireless channels \cite{Arun1}. It is shown that under a wide range of parameters our adaptive algorithm improves the throughput and delay performance of the IEEE 802.11 DCF.

MPR MAC protocols supporting QoS (Quality of Service) provisioning have been proposed in the past. A protocol named Multi Reservation Multiple Access (MRMA) was proposed by Hui Chen et al.\cite{Chen2005a}. The authors propose a centralized reservation scheme for channel access which provides QoS for multimedia traffic. The upcoming Wireless LAN standard 802.11ac, which supports optional MU-MIMO \cite{6687187}, an MPR enabling technology, uses Enhanced Distributed Channel Access (EDCA) for medium access. EDCAF (EDCA Function) is an extension of DCF to support priority traffic.

In this paper, we propose an enhancement to the IEEE 802.11ac EDCA protocol for channels supporting MPR for QoS provisioning.  The modified protocol incorporates an adaptive backoff mechanism that decrements the backoff counter value  as a function of number of ongoing transmissions in the channel. We show that in the case of MPR channels, in addition to $CW_{min}$, $CW_{max}$, and $AIFSN$, we can use two more parameters namely (i) threshold and (ii) Counter decrementation value, that can offer service differentiation.
Our protocol operates in a fully distributed fashion and do not require any form of centralized coordination. 
  
  The remainder of the paper is organized as follows. Section \ref{sec:model} describes the MPR channel and network models. In Section \ref{sec:protocoldesign}, we briefly review the adaptive backoff scheme. The proposed enhancements to the 802.11ac for MPR channels is discussed in section \ref{sec:servicediff}. In Section \ref{sec:perfeval}, the simulation setup and the performance evaluation of the proposed  protocol is presented. Section \ref{sec:conclusion} concludes the paper.

\section{System Model} \label{sec:model}
\subsection{Channel Models}
The MPR channel models which are widely used in literature include the $k$- MPR channel model and the generalized MPR channel model.
In a $k$ - MPR channel, as long as the number of packets transmitted is not greater than $k$, a node will be able to receive all packets without loss.
In case the number of transmissions exceeds $k$, collision occurs and nodes will not be able to receive any of the packets. Suppose $\zeta$ denotes the number of concurrent transmissions in a collision domain,
\[\Pr(\mbox{Success}) = \left\{
\begin{array}{l l}
  1 & \quad \mbox{if $\zeta \le k$}\\
  0 & \quad \mbox{if $\zeta >  k$}\\
\end{array} \right. \]
In nutshell, in a $k$-MPR channel, either all transmissions are successful or none of them are successful. Such a case  occurs when the probability of successful reception directly depends on the interference level at the receiver (SINR).
In more generalized MPR channel due to Ghez et al. \cite{ghez1988}, a node will be able to receive  $m$ out of $n$ transmissions ($m$ $\leq$ $n$) with certain nonzero probability. However we adopt the simpler $k$-MPR channel in our work. 

\subsection{Network Model}
We consider an ad-hoc wireless network in which nodes operates in a distributed manner. Every node is equipped with receivers capable of receiving up to $k$ transmissions concurrently without error. Nodes are assumed to be half-duplex; i.e. it is not possible for a node to transmit and receive simultaneously. We further assume that all nodes can perform  enhanced carrier sensing, i.e. non-transmitting nodes have the capability to estimate accurately the number of ongoing transmissions.
\section{IEEE 802.11 DCF based Adaptive backoff algorithm} \label{sec:protocoldesign}
In this section, we briefly explain the adaptive backoff algorithm \cite{Arun1}, which is required to understand the proposed enhancement to IEEE 802.11ac protocol for QoS provisioning under MPR. 
The basis of the IEEE 802.11 DCF protocol is the CSMA/CA (Carrier Sense Multiple Access/Collision Avoidance) scheme with binary exponential backoff \cite{6178212}.
The channel is assumed to be slotted.
In 802.11 DCF, a station having a packet to transmit senses the medium.  If the channel remains idle  for a duration equal to the DIFS, the station proceeds with transmission.
If the medium is sensed to be busy, the station waits for the current transmission to get over. 
The station then generates a random backoff value drawn uniformly from the interval $0$ to $CW_{min}$ (minimum contention window size) and continues to sense the channel. This backoff count is decremented at the end of each idle slot and frozen whenever the channel becomes busy. The node attempts a transmission when the backoff count reaches zero. If the transmission is successful (receipt of ACK frame), the next packet is processed. On the other hand if the packet transmission is unsuccessful, the contention window is doubled and the random backoff process begins. The process is continued until the transmission is successful or until maximum number of retries is reached upon which the packet is dropped.

 In the adaptive backoff protocol, the node uses enhanced carrier sensing to estimate the number of ongoing transmission. The backoff counter is frozen only when the number of ongoing transmissions is greater than or equal to a threshold. The value of the threshold can be fixed to be equal to or less than the MPR capability of the node. Further, the backoff counter will be decremented by the number of additional possible transmissions. Suppose the MPR capability of the node is $K$ and there are $i$ ongoing transmissions. The adaptive backoff protocol decrements the counter by $K-i$. If we denote $d(i)$ as the amount by which the backoff counter is decremented when a slot time elapses in which $i$ transmissions are going on, for the adaptive backoff protocol,
\[d(i) = \left\{
\begin{array}{l l}
  K-i & \quad \mbox{$i \le K_t$}\\
  0 & \quad \mbox{otherwise}\\
\end{array} \right. \]
where $K_t$ ($<K$) is the threshold and $K$ is the MPR limit.

The protocol adapts itself to the traffic conditions. If the number of transmission is small as compared to the MPR capability then the counter gets decremented faster leading to lesser delay and more throughput.

We retain the definitions of SIFS (Short Inter Frame Space), DIFS (DCF Inter Frame Space), EIFS (Extended Inter Frame Space) as specified in the IEEE 802.11 standard \cite{6178212}. However from the MAC layer's perspective the channel becomes Idle or Busy under the following condition. An "Idle Slot" is one in which number of ongoing transmissions is less than or equal to the threshold $K_t$. A slot is defined to be "Busy" only if the number of transmission exceeds the threshold $K_t$. Since the adaptive protocol can decrement  the counter by a number greater than one, the counter may reach even negative values without ever reaching zero. This calls for redefining the condition for transmission. The nodes should attempt a transmission as soon as the counter reaches a nonpositive integer value. When a node attempts a transmission, it checks whether the  channel continues to remain idle for a duration of DIFS in which no more than $K_t$ transmissions take place. 

An important consequence of not freezing the counter during an ongoing transmission is that the transmissions from different nodes may not be frame synchronous. Further, an ongoing transmissions can encounter collision any time during its transmission for the following reason:
When the number of ongoing transmissions, is less than or equal to threshold, two or more nodes can continue to count down and reach a nonpositive counter value at the same slot. As a result, they may begin their transmissions. If the resultant number of transmissions (ongoing plus the newly initiated transmissions) exceeds the MPR limit, collision occurs, not only to the newly initiated transmission but also for the already ongoing transmissions.
  In other words, a transmission can be declared to be successful only if it does not encounter collision until its completion.
\section{Service differentiation for MPR with Enhanced carrier sensing}
\label{sec:servicediff}
In this section we propose an enhancement to the performance of EDCA protocol of 802.11ac with the use of adaptive backoff protocol. The basic idea is to set higher thresholds for high priority traffic and lower thresholds for lower priority traffic. Additional service differentiation is obtained by adopting different algorithms for counter decrements based on the priority of the traffic i.e. adaptive algorithm for higher priority traffic and non adaptive algorithm for low priority traffic.
 In order to validate the use of $K_t$ as a service differentiation parameter, we have carried out the simulation for four different access categories which roughly corresponds to urgent, real time and non real time data traffic as shown in Table \ref{table:}.
\begin{table}[!t]
\caption {Access Categories and intended applications} 
\centering 
\begin{tabular}{ | c | c | c |}
\hline 
\textbf{{Access category}} &\textbf{{Application}}\\ [1 ex] 
\hline\hline 
Access category 0 ($AC_{0}$) & High priority, real time Voice  \\ \hline
Access category 1 ($AC_{1}$) & High priority, Video play back  \\ \hline
Access category 2 ($AC_{2}$) & Medium priority, Best Effort  \\ \hline
Access category 3 ($AC_{3}$) & Lowest priority, Back ground, file transfer \\   
\hline 
\end{tabular}
\label{table:} 
\end{table}
In order to differentiate the Quality of Service (QoS) given to each access categories, we set different values of thresholds for different Access Categories(ACs). The use of adaptive / non-adaptive counter decrementing strategies are also used for service differentiation. Table \ref{tab:acdiff} summarizes the access categories and service differentiation techniques, where $L$ corresponds to the estimated number of transmissions. $K_{t}$ is the threshold and $K$ is the MPR limit.
\begin{center}
\begin{table*}[!ht]
\caption[Access categories and parameters for service differentiation]{Access categories and service differentiation \label{tab:acdiff}}
\normalsize
\centering
\begin{tabular}{|c|p{10.5cm}|}
\toprule
Access Category & Parameters \\
\midrule 
AC$_{0}$ &  Kt = K-1, and count down is adaptive = K - L\\
AC$_{1}$ & Kt = $\lceil{ K/2 }\rceil$ and count down is adaptive = K - L\\
AC$_{2}$ &  Kt = $\lceil {K/4}\rceil$ and count down is non adaptive, always decrement by 1.\\
AC$_{3}$ & Kt = 1 and count down is non adaptive, always decrement by 1. \\
\bottomrule 
\end{tabular}
\end{table*}
\end{center}
\vspace*{-1cm}
\section{Performance Evaluation} \label{sec:perfeval}
\subsection{Simulation Setup}
The network model adopted in our simulation corresponds to  $N$ users with uplink to a common access point. In other words  $N$ nodes are communicating with a central node having MPR capability of $K$, through a common wireless channel. This situation is equivalent to an ad-hoc network of $N$ stations, each having MPR capability of $K$, where data exchange takes place between two arbitrary stations.
The packet arrival processes at each node is a Poisson process independent of arrivals at other nodes. Further, all packet lengths are of fixed size. We assume $k$-MPR channel model for the simulations although the proposed protocol can be used under generalized MPR channels. We assume ideal channel conditions - transmission errors occur only as a result of collisions. The simulations were carried out for basic access only (no RTS/CTS).

In our simulation, we have used the network parameters given in Table \ref{tab_simparams}, mostly taken from IEEE 802.11 standard \cite{6178212} for FH-PHY. The simulations are done using SimPy \cite{SimPy} discrete event simulator, using Python. Since our goal was to study the MAC layer performance of different protocols, the details of the PHY layer were omitted.
Therefore a simulator with an ideal physical layer was implemented using SimPy discrete event simulator.  
The throughput, MAC delay and jitter for each access categories were computed. The results are described in the next section.
\begin{table}[!t]
\renewcommand{\arraystretch}{1}
\caption{Simulation Parameters \label{tab_simparams}}
\centering
\begin{tabular}{ l l }
\hline
    \hline
  Parameter & Value \\
  \hline
  Packet payload & 8184  bits \\
  MAC Header & 272 bits \\
  PHY Header & 128 bits \\
  Channel Bit Rate & 1Mbps \\
  Slot time ($\sigma$) & 50 $\mu s$ \\
  DIFS & 128  $\mu s$ \\
  Max backoff stage ($m$) & 5 \\
  Retry limit  & 4 \\
  \hline
    \hline
\end{tabular}
\end{table}
\subsection{Results}
\begin{figure}[!t]
\centering
\includegraphics[width=0.9\columnwidth]{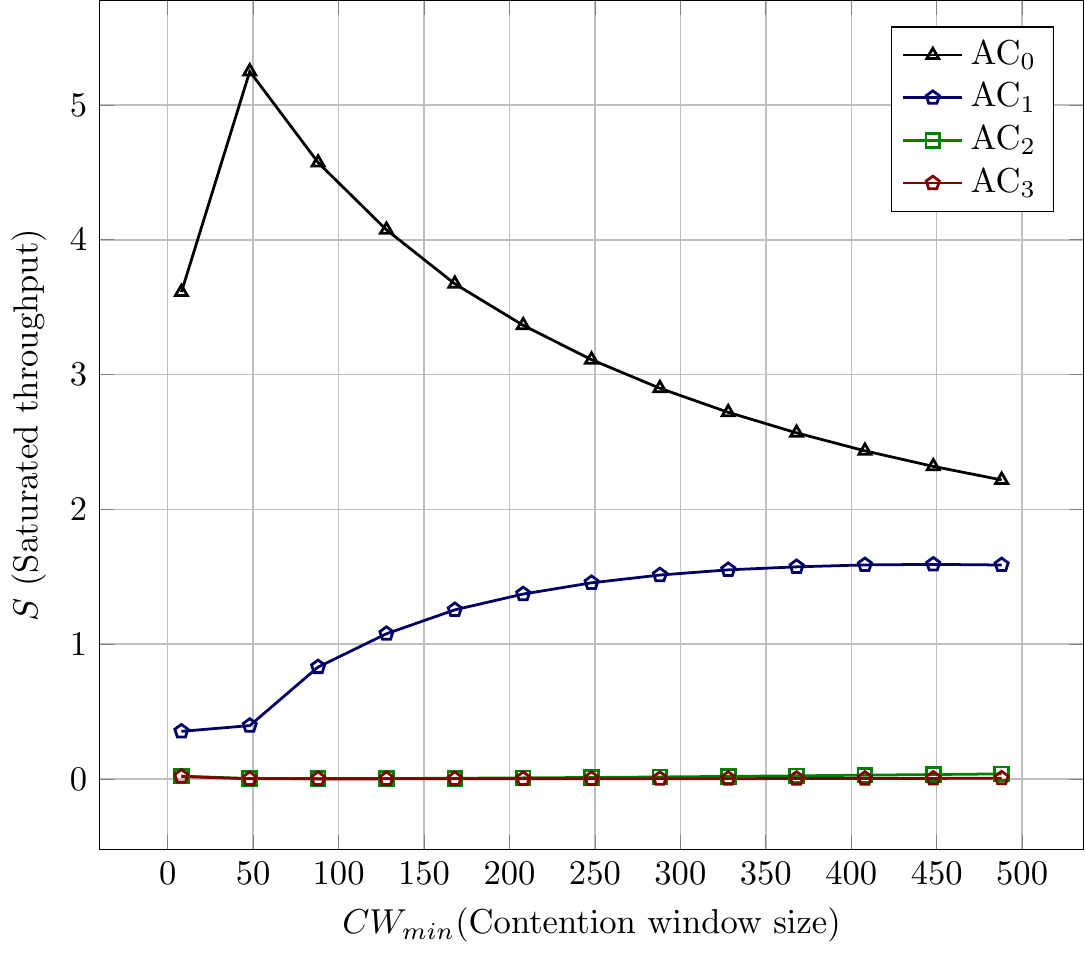}
\caption{The saturated throughput against contention window size $CW_{min}$ for different access categories (\emph{Params}: Number of stations $N = 40$, $K=8$)}
\label{fig:sattpqos_cw8}
\vspace*{-4mm}
\end{figure}

\begin{figure}[!t]
\centering
\includegraphics[width=0.9\columnwidth]{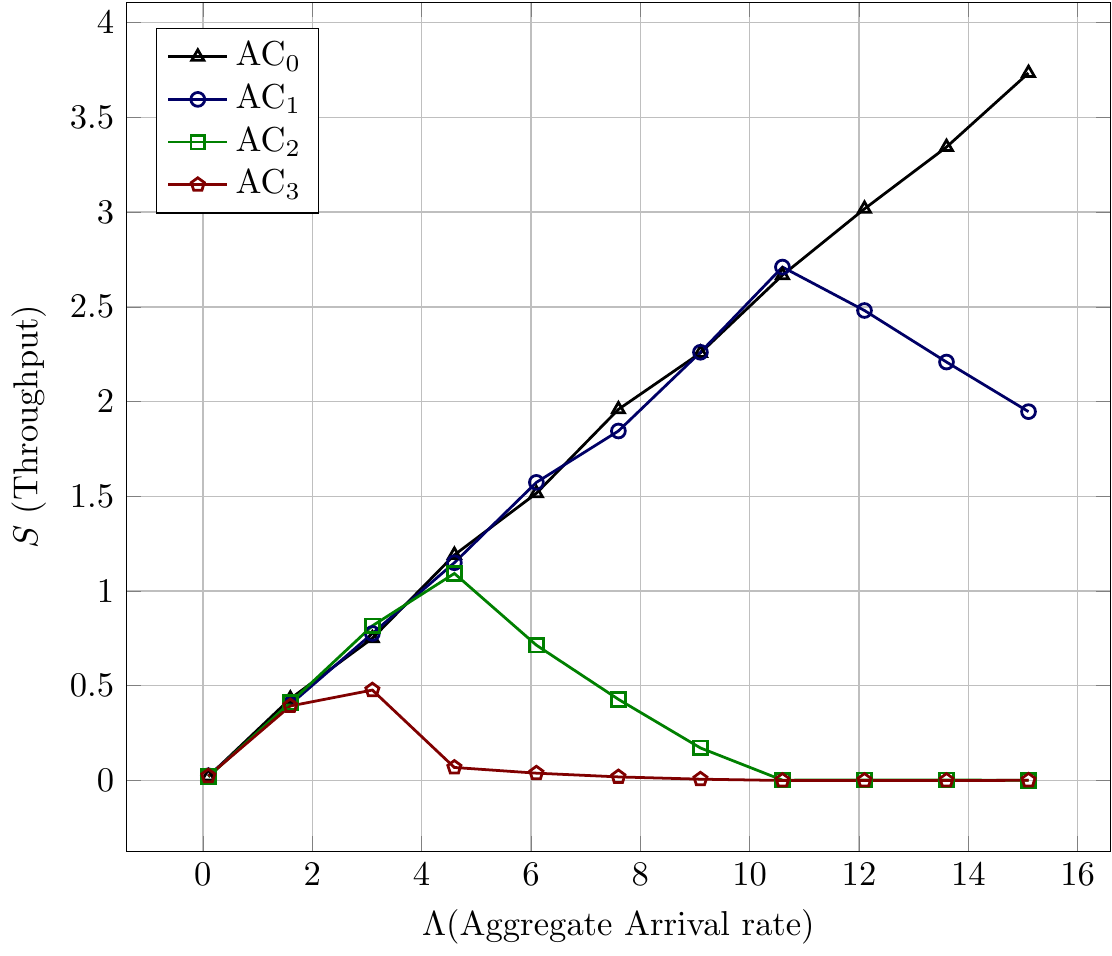}
\caption{The throughput against offered traffic for different access categories (\emph{Params}: Number of stations $N = 40$, $K=8$, $m=7$, $CW_{min}=256$)}
\label{fig:unsattpqos_offered8}
\end{figure}

We first investigate  the effect of contention window size for different access categories by varying the $CW_{min}$. 
In Fig. \ref{fig:sattpqos_cw8}, the saturation throughput is plotted against $CW_{min}$. The high priority traffic gets higher share of the channel bandwidth as expected. It is to be noted that the service differentiation is maximum when $CW_{min}$ is set at a value of $50$ much greater \cite{Arun1} than what is recommended by the EDCA standard.    However to keep the delay small, the $CW_{min}$ can be set to a lower value such as $16$. There is a trade off between service differentiation and the acceptable delay of $AC_{0}$. In our simulation the $CW_{min}$ is varied till $500$ to study the variation of saturation throughput with changes in contention window size. Note that both $AC_{0}$ and $AC_{1}$ use adaptive protocol but with different thresholds. It is interesting to note that for large $CW_{min}$, the performance of $AC_{1}$ and $AC_{0}$ approach each other. In order to report this finding we retained the plot till $CW_{min}$ = 500. The two plots shows very less throughput for $AC_{2}$ and $AC_3$ since the network is operating in saturation. In such a case, The channel is filled with high priority traffic and low priority traffic is not carried at all. From a node's point of view, since the channel is seldom idle, the counters for low priority traffic do not get decremented often.
In Fig. \ref{fig:unsattpqos_offered8}, the throughput is plotted against normalized offered traffic. At very small arrival rates, all the access categories get access to channel and almost all traffic are carried. When the network load increases, the priorities come into picture. At moderate loads, the low priority traffic is effectively excluded from channel access. At still higher traffics, only high priority packets are being transmitted. As the offered traffic is increased the number of concurrent transmissions also increases. Low priority access categories having lower threshold ($K_{t}$) would freeze their back off counter and defer the transmission attempts. This results in higher throughput for higher priority access categories. 
\begin{figure}[!t]
\centering
\includegraphics[width=0.9\columnwidth]{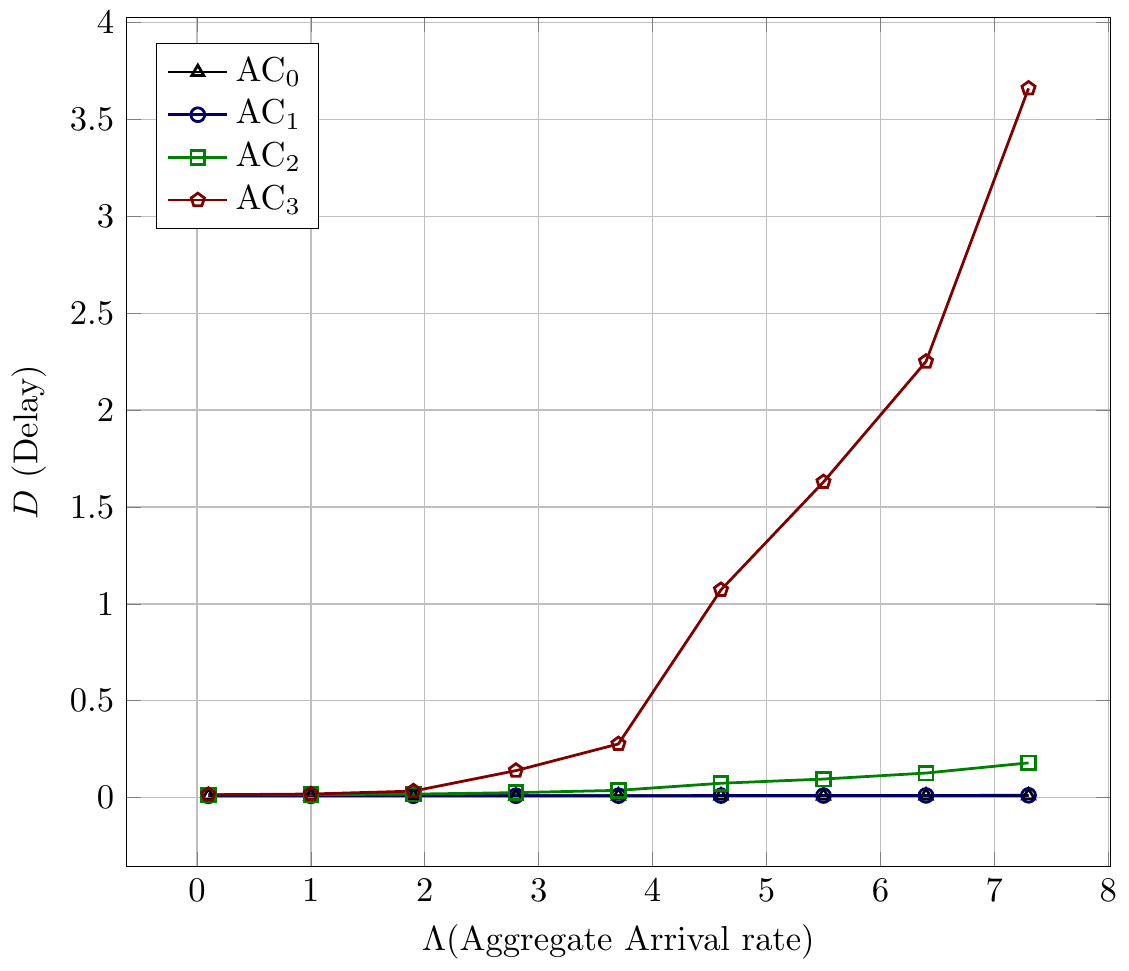}
\caption{MAC delay against offered traffic for different access categories (\emph{Params}: Number of stations $N = 40$, $K=8$)}
\label{fig:unsatdelayqos_offered8}
\vspace*{-5mm}
\end{figure}
In  Fig. \ref{fig:unsatdelayqos_offered8}, the MAC delay is plotted against normalized offered traffic or aggregate arrival rate. The offered traffic is normalized with respect to the packet transmission time.  Since the MPR channel can support more than one transmissions normalized  arrival rate of more than 1 is allowed. The MAC delay plot in Fig. \ref{fig:unsatdelayqos_offered8} follows a similar pattern as we found for throughput. At lower arrival rates, the channel is mostly idle and most of the packets are transmitted right away. Therefore the MAC delay is close to zero for all ACs. When the offered traffic is increased, the expected number of ongoing transmission increases. Lower priority access categories having lower threshold freeze their backoff counter more often leading to larger delay. However higher priority access categories continue  their countdown leading to lesser delay. Further due to the  adaptive nature of countdown process the mean value of the counter decrement ($K-L$) is more for $AC_{0}$ as compared to $AC_{1}$. As a result even at very large offered traffic the delay of highest priority access category is kept very low.

\begin{figure}[!t]
\centering
\includegraphics[width=0.9\columnwidth]{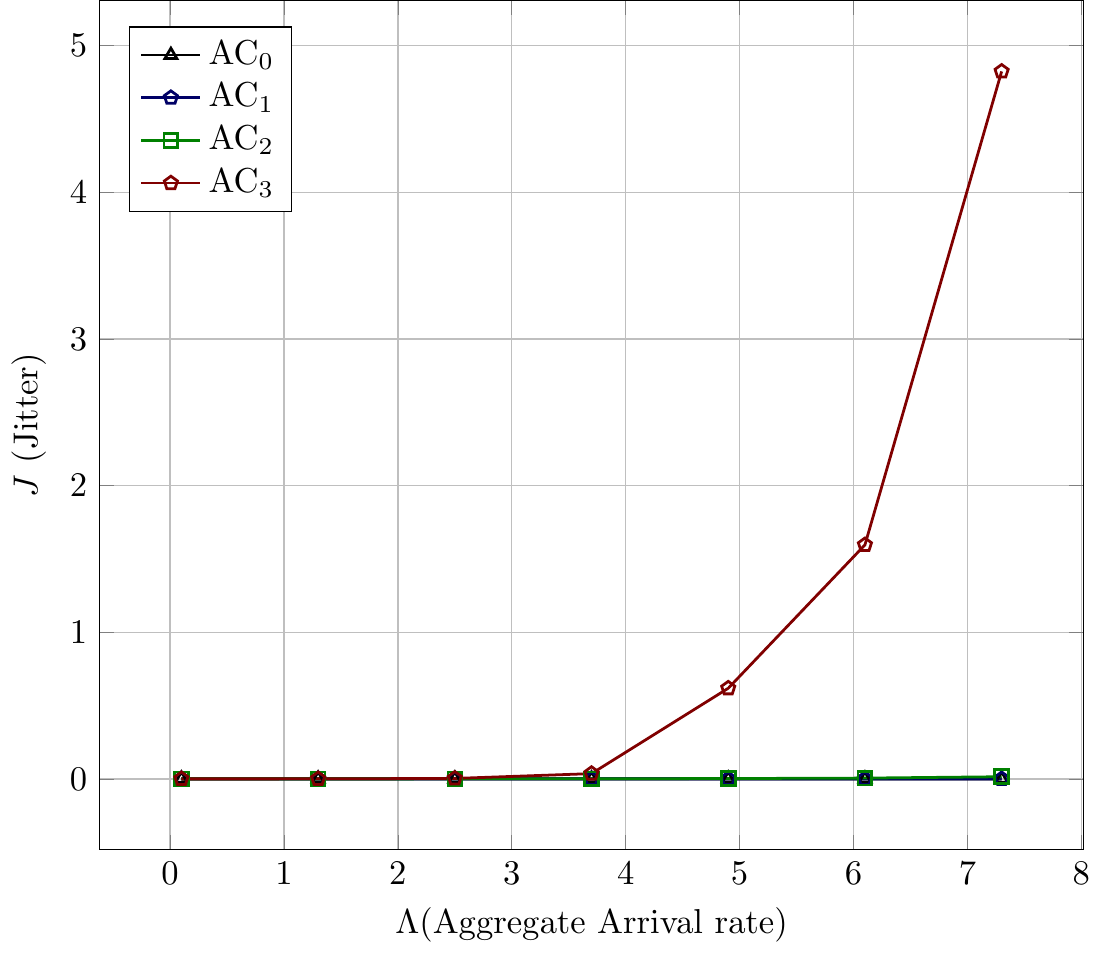}
\caption[Jitter for different access categories]{Jitter against normalized offered traffic for different access categories (\emph{Params}: Number of nodes $N = 40$, $K=8$)}
\label{fig:unsatjitterqos_offered8}
\vspace*{-2.5mm}
\end{figure}
In Fig. \ref{fig:unsatjitterqos_offered8}, the jitter (packet delay variation) is plotted against offered traffic. We use the variance of the MAC delay to compute jitter. The jitter closely follows the delay curve. Whenever the number of ongoing transmissions changes around $\lceil{ K/2 }\rceil$ the $AC_{1}$ switches between countdown and freezing of the backoff counter.  A similar situation occurs around $\lceil{ K/4 }\rceil$  and $K_{t}$=1 for $AC_{2}$ and $AC_{3}$ respectively. This switching between countdown and freezing of the backoff counter gives rise to larger jitter in low priority access categories.  $AC_{0}$ rarely has to freeze its backoff counter value leading to lesser jitter. 
\section{Conclusion}\label{sec:conclusion}
In this paper, we have proposed an enhancement to the 802.11ac EDCA protocol using an adaptive backoff procedure. This protocol  does not require any additional memory or computations. Yet with this simple design, the adaptive protocol achieves significant performance improvement for high priority traffic leading to QoS provisioning.
 In this paper, we have made a crucial assumption that the stations are able to accurately determine the number of ongoing transmissions using enhanced carrier sensing. In reality this estimate may vary from the actual number of transmissions. In spite of such errors in the estimated number of transmissions, it can be seen that our proposed adaptive backoff protocol will  still work although suboptimally.
Future work includes the theoretical performance analysis of the proposed protocol using stochastic models and  analyzing the effect of error in the estimate of enhance carrier sensing on the performance of the protocol.






%




\bibliographystyle{IEEEtran}
\bibliography{citedin}

\end{document}